\documentclass[preprint]{aastex}

\usepackage{graphicx}
\usepackage{wrapfig}
\usepackage{natbib}

\newcommand{\be}{\begin{equation}}
\newcommand{\ee}{\end{equation}}
\newcommand{\nn}{\mbox{} \nonumber \\ \mbox{} }
\newcommand{\ba}{\begin{eqnarray}}
\newcommand{\ea}{\end{eqnarray}}

\newcommand{\Alfven}{Alfv\'{e}n }

\newcommand\eg{{\it{e.g.\ }}}

\newcommand{\Bf}{{magnetic field}}

\newcommand{\Ef}{{electric  field}}

\newcommand{\NS}{neutron star}

\newcommand{\EM}{electromagnetic}

\newcommand{\Sc}{Schwarzschild}
\newcommand{\ms}{magnetosphere}
\newcommand{\mss}{magnetospheres}

\newcommand{\LC}{light cylinder}
\newcommand{\Lf}{Lorentz factor}

\begin{document}

\title{Magnetic  loading of magnetars' flares}
\author{Maxim Lyutikov}
\affil{Department of Physics  and Astronomy, Purdue University,   525 Northwestern Avenue, West Lafayette, IN47907-2036, USA; lyutikov@purdue.edu}

\begin{abstract}
Magnetars,  the likely  sources   of Fast Radio Bursts (FRBs),  produce both  steady highly relativistic  magnetized  winds, and occasional    ejection events.
We demonstrate that the requirement of conservation of the  magnetic flux dominates  the overall dynamics of magnetic  explosions.   
This is missed in conventional hydrodynamic models of the ejections as expanding shell with parametrically  added \Bf,  as well as   one-dimensional models of magnetic disturbances.  Most of the initial  free energy  of an explosion  is  actually spent on stretching its   own  internal \Bf, while  doing  minimal  $pdV$ work against the surrounding. Magnetic explosions from magnetars come into force balance with the  pre-flare wind close to the \LC.  They are then  advected quietly  with the wind, or propagate as electromagnetic disturbances. No  powerful shock waves  are generated in the wind.
\end{abstract}

\maketitle

\section{Introduction}

Observations of  correlated radio and X-ray bursts \citep{2020Natur.587...54C,2020arXiv200511178R,2020Natur.587...59B,2020ApJ...898L..29M,2021NatAs.tmp...54L} 
established the  FRB-magnetar connection.

Two kind of models of FRBs' {\it loci}  are competing at the moment. The first advocates that FRBs are   magnetospheric  events, \eg\ Solar flare-like  \citep{2002ApJ...580L..65L,2006APS..APR.X3003L,2006MNRAS.367.1594L,2020arXiv200505093L,2020ApJ...897....1L};  the emission mechanism remains unidentified, akin to the 50+ years problem of pulsar radio emission \citep[see][for new ideas]{2020ApJ...897....1L,2021arXiv210207010L}.
Second are   wind-generated GRB-like events \citep{2014MNRAS.442L...9L,2017ApJ...843L..26B,2019MNRAS.485.4091M, 2017ApJ...841...14M,2019MNRAS.485.4091M,2020ApJ...896..142B}. Emission mechanism is the cyclotron maser \citep{1994ApJ...435..230G,2019MNRAS.485.3816P,2020MNRAS.tmp.2086B}. 
\citep[For discussion of general constraints on plasma parameters and models see \eg][]{2019arXiv190103260L,2021Univ....7...56L}

In this paper we argue that wind-model of FBRs are internally inconsistent. Qualitatively, these models use the hydrodynamic paradigm of the internal shocks models of GRBs, the  flying shells \citep{2004RvMP...76.1143P}, with parametrically  added   \Bf\ component. 
This simple addition of \Bf\ cannot be applied  in principle to the magnetic explosions of magnetars. The key point  is the conservation of the magnetic flux  {\it within}  the exploding plasma. This is related 
to the  sigma-problem in pulsar winds \citep{1974MNRAS.167....1R,1984ApJ...283..694K},   reformulated  by \cite{2003astro.ph.12347L,2006NJPh....8..119L}
as a magnetic flux conservation problem. 
The theoretical difference  between the two model  (the  ``magnetized shells''  and the  present model),  are enormous.  Instead of highly relativistic shock with the \Lf\ over a million, as advocated by  \eg\ \cite{2020ApJ...896..142B}, the magnetic explosions comes into a force balance approximately near the \LC, and propagates as an \EM\ pulse there on. No powerful shocks are generated. Hence no emission of FRBs from the far wind.

We first give an  analysis  of  the  wind-FRB models  \citep{2019MNRAS.485.4091M,2020ApJ...896..142B} from the point of basic theory  of  pulsar winds and relativistic shock propagating through the winds, \S \ref{prelim}.
These are the types of ``double relativistic explosion'' previously considered in various set-up by 
\cite{2002PhFl...14..963L,2011MNRAS.411..422L,doi:10.1063/1.4977445,2017ApJ...835..206L,2021ApJ...907..109B}.
In the main \S \ref{Magnetically-driven} we argue that an effective ``magnetic loading'' quickly reduces the power of the magnetar's  explosion, producing either an EM pulse through the wind, or a confined magnetic structure in pressure balance with the wind. No ultra-relativistic shocks are produced. 


\section{The  pulsar wind and  shock dynamics  in the magnetized wind}
\label{prelim}

\subsection{Pulsar wind primer}

Pulsar and magnetars produce relativistic highly magnetized winds \citep{1969ApJ...158..727M,1970ApJ...160..971G,1973ApJ...180L.133M}, that can be parametrized by the wind luminosity $L_w$ and the ratio of Poynting  to particle fluxes:
\be
\mu_w =\frac{L_w}{ \dot{N}  m_e c^2} =  \frac{B_{LC}^2}{4\pi n_{LC} m_e c^2} 
\ee
where $\dot{N}$ is the rate of lepton ejection by the \NS; subscript LC indicates values measured at the \LC,
\be
R_{LC} = 4\times 10^9 \, P\, {\rm cm}
\ee
for the period of a star  $P$ measured in seconds. 
Expected values of $\mu_w$ are in the range of $10^4-10^6$ \citep{AronsScharlemann,arons_83,Hibschman,2007ApJ...657..967B}. In the case of GJ scaling of density, with multiplicity $\kappa = n/n_{GJ}$,
\be
\dot{N} = \kappa \frac{ \sqrt{ c L_w}}{e} 
\ee

Initially the wind accelerates linearly away from the \LC\
\be
\Gamma_w = \frac{r}{R_{LC}}
\ee
so that the  wind magnetization decreases
\ba &&
n' =\left( \frac{R_{LC}}{r} \right)^{-3} n_{LC} 
\nn && 
B' = \left( \frac{R_{LC}}{r} \right)^{-2} B_{LC}
\nn &&
\sigma(r) = \frac{B^{\prime,2}}{4 \pi n' m_e c^2} =  \frac{R_{LC}}{r} \mu_w
\label{sigma1}
\ea
(primes denote quantities measured in the wind frame.)
\footnote{ \label{ft} This definition of magnetization parameter $\sigma_w $ is relevant only at $r\gg R_{LC}$ as it does not take into the account large parallel momenta of particles near the \LC.}

The terminal value for  the wind acceleration is the \Alfven\ condition,
\ba  && 
\Gamma_w = \sigma_w^{1/2}=  \mu_w ^{1/3} 
\nn &&
 \sigma_w =  \mu_w ^{2/3} \gg 1 
\ea
reached at 
\be
 R_w =  \mu_w ^{1/3}R_{LC} = \Gamma_w R_{LC}
\label{rw}
\ee

Total energy and mass (in lab frame) contained in the  acceleration region, 
\ba &&
E_w = L_{w} \frac{R_w} {c} 
\nn &&
M_w = m_e \dot{N} \frac{R_w} {c} 
\ea
are typically insignificant; mass loading of the following  shock is further reduced by $1/\Gamma_w$. 

At distances $r>  R_w$
\ba &&
 \Gamma_w = \mu_w^{1/3} 
  \nn &&
n' =\mu ^{-1/3} \left( \frac{R_{LC}}{r} \right)^{-2} n_{LC} 
\nn && 
B' =  \mu ^{-1/3} \left( \frac{R_{LC}}{r} \right)^{-1}  B_{LC}
\nn &&
\sigma_w=  \mu_w^{2/3} 
\nn &&
\Gamma^{\ast}= 2 \sqrt{\sigma_w} \Gamma_w  =2 \mu_w^{2/3}= 2 \Gamma_w^2
\label{kw2}
\ea
Quantity $\Gamma^\ast $ is the  \Lf\  (measured in lab frame) required to produce a shock in the receding wind.

The above  considerations assumes  no magnetic dissipation. At the basic level it is not  consistent with observations of the PNWe.
As \cite{1974MNRAS.167....1R,1984ApJ...283..694K}  argued,  $\sigma_w$ cannot  remain $\gg 1$ until the  (conventional MHD)  wind termination shock. It should drop to $\leq 10^{-2}$. This can happen either close to the LC \cite{lyubarsky_kirk_01,1990ApJ...349..538C} or at the termination shock \citep{LyubarskyStripedWind,2009ApJ...698.1523S,2011ApJ...741...39S}. As   \cite{2014MNRAS.438..278P} argued, magnetic dissipation  can occur  in the bulk of the nebula \citep{2017SSRv..207..137P}; still observations of the Crab's Inner knot  require that a large sector of the wind has  low magnetization \citep{2016MNRAS.456..286L}.
In the limit of dominant dissipation in the near wind, $\sigma_w \leq 1$, all of spin-down is carried by particle flow, hence in that case $\Gamma_w = \mu_w \sim 10^4-10^6$.


\subsection{Explosions  in relativistic  winds}

Let us next  consider explosions in the preceding magnetar wind. We assume that some kind of the central engine (a magnetar) produces energetic events on top of the steady wind. 
This will constitute a type of double relativistic explosion \citep{2002PhFl...14..963L,2011MNRAS.411..422L,doi:10.1063/1.4977445,2017ApJ...835..206L,2021ApJ...907..109B}. First, we solve a formal MHD problem, and then discuss its limitations.

\subsubsection{Point explosion in relativistic fluid wind}
Consider a relativistic  fluid wind of luminosity $L_w$  moving with \Lf\ $\Gamma_w$
\be
L_w = 4\pi r^2 \Gamma_w^2 n_w' m_e c^3 = \Gamma_w \dot{M}_w c^2
\ee
where $n_w'$ is density in the wind frame; the lab frame density is $n_w= \Gamma_w n_w'$. 

Consider an   explosion that involves energy  $E_{ej}$ and no mass $M_{ej}=0$. Thus, we assume that the energy of the explosion is transferred to the wind instantaneously. 
Consider a shock moving with \Lf\ $\Gamma \gg \Gamma_w$  (as measured in lab frame). In a shock frame typical energy of post-shock particles is
$T_2 \sim \Gamma/(2\Gamma_w)$. The shock sweeps particles giving them bulk \Lf\ $\Gamma$; at radius $r$ the swept-up mass (in lab frame) is $L_w r/(c^2 \Gamma_w)$. Thus
\be
\Gamma \sim \sqrt{2}  \Gamma_w \sqrt{ \frac{R_0}{r}}
\label{Gammafluid}
\ee
where 
\be
R_0=
\frac{c E_{ej}}{L_{w}}
\label{R00}
\ee

\subsubsection{Point explosion in relativistic magnetized wind}

Relativistic explosions in static magnetic configurations were studied by \cite{2002PhFl...14..963L}, producing  self-similar solutions solutions of the kind of \cite{BlandfordMcKee}.  Let us generalize them to the moving media. Below we are not interested in the structure of the flow  \cite[it will be the same as found in ][]{2002PhFl...14..963L}, but in the overall scaling of the shock \Lf\ $\Gamma$ dependance  on the wind parameters. 

Let us first consider a simple, extreme case of point magnetospheric  explosion with no mass loading. Let the  wind preceding the ejection have   magnetization $\sigma_w \gg 1$. Neglecting small contribution to the wind luminosity from the particles
\be
L_w = 4\pi r^2 \Gamma_w^2 \frac{ B_w^{\prime,2}}{4\pi}  c 
\ee

Consider again  an   explosion that involves energy  $E_{ej}$ and no mass $M_{ej}=0$.
The   interaction in the accelerating region $r\leq R_w$, before the wind reaches  $\Gamma_w$, does not affect much the flow,  because of small $E_w$ and $M_w$, high values of $\sigma(r)$ and the fact that both the ejecta and the wind accelerate linearly with $r$. (This statement has been verified  according to the  following discussion.)


Consider a shock moving with \Lf\ $\Gamma$  (as measured in lab frame) through  a highly magnetized wind, which itself is moving with $\Gamma_w$. 
Let the values of the \Bf\
 in the lab frame before the shock be $B_w$,
 \ba && 
 B_w = \frac{\sqrt{L_{w}}}{\sqrt{r} c}
 \nn &&
 B_w' = \frac{B_w}{\Gamma_w}
 \ea
 (prime is in wind frame).
 The shock is moving through the wind with 
 \be
 \Gamma_s = \frac{\Gamma}{2\Gamma_w}
 \ee
 In the frame of the shock the shocked part of the wind moves with \Lf\ $\sqrt{\sigma_w}$.  It carries \Bf\
 \be
 B_2' = \frac{\Gamma}{2 \Gamma_w \sqrt{\sigma_w}} B_w'  
 \label{B2prime}
 \ee
 as measured in the post-shock frame. The post shock frame moves with \Lf\ $\Gamma/( 2 \sqrt{ \sigma_w})$ in lab frame, 
 thus
 \be
 B_{w,2}= \left(  \frac{\Gamma}{\Gamma^{\ast}}\right)^2 B_w
 \label{Bpost}
 \ee
 This is \Bf\ in the post shock region as measured in the lab frame.
 
 The swept-up material is located within 
 \be
 \Delta r \sim \frac{\sigma}{\Gamma^2} r
 \ee
 
 The energy budget reads (particle contribution is neglected for $\sigma_w \gg 1$)
 \be
  E_{ej} =c   \left({B_{w,2}^2}  \right) r^2 \Delta r\approx {L_w r} \frac{\Gamma^2}{4 \sigma_w \Gamma_w^4}
  \ee
  
  Thus,
  \be
  \Gamma= 2  \Gamma_w^2 \sqrt{\sigma_w}   \sqrt{  \frac{R_0}{r}} \to  2  \sqrt{  \frac{R_0}{r}}   \Gamma_w^3
  \label{GammaR}
  \ee
  where the last relation assumed $ \Gamma_w = \sqrt{\sigma_w}$.
  Notice the difference in the power of $\Gamma_w$ in comparison with the fluid case (\ref{Gammafluid}).

In observer time
\ba &&
t_{ob} = \frac{t}{2  \Gamma^2} = \frac{ c t^2}{8 R_0 \Gamma_w^4 \sigma_w}
\nn &&
 \Gamma = 2^{1/4}   \Gamma_w  \sigma_w^{1/4}  \left( \frac{R_0} { c t_{ob}} \right)^{1/4} 
 \ea

The energy is concentrated in 
\be
\frac{\Delta r}{r} = \frac{1}{8 \Gamma_w^4} \frac{r}{R_0} 
\ee
(independent of $\sigma_w$)

\subsection{Shock interaction of  relativistic  winds}

Let's  next assume that the first wind is followed by the second  wind from the flare with \Lf\ $\Gamma_f$,  wind frame \Bf\ $B_f$ and magnetization $\sigma_f$
\be
L_f = \Gamma_f^2 B_f^2 r^2 c
\ee

Assume strong interaction, so that a reverse shock  (RS) is generated in the flare wind and  forward shock (FS)  is generated in pre-explosion wind. Let in the lab frame the contact discontinuity between two winds move with \Lf\ $\Gamma_{CD}$. Then
\ba &&
\Gamma_{FS} = 2 \sqrt{\sigma_w} \Gamma_{CD}
\nn &&
\Gamma_{RS} =  \frac{\Gamma_{CD}}{ 2 \sqrt{\sigma_f}}
\ea
Lorentz factor of the RS with respect to the flare flow is
\be
\Gamma_{RS}' = \frac{\Gamma_f}{\Gamma_{CD} \sqrt{ \sigma_f}}
\ee

Balancing \Bf\ in the shocked wind and shocked flare flows
\be
\frac{\Gamma_{CD} }{\Gamma_w^2} \frac{\sqrt{L_w}}{\sqrt{c} r}  =\frac{1}{2 \Gamma_CD}    \frac{\sqrt{L_f}}{\sqrt{c} r},
\ee
we find
\ba && 
\Gamma_{CD}=  \left( \frac{L_f}{L_w}\right)^{1/4}  \Gamma_w
\nn &&
\Gamma_{FS} = 2 \left( \frac{L_f}{L_w}\right)^{1/4}  \Gamma_w \sqrt{\sigma_w}
\nn &&
\Gamma_{RS} = \frac{1}{2}  \left( \frac{L_f}{L_w}\right)^{1/4}  \frac{ \Gamma_w} {\sqrt{\sigma_f}}
\nn &&
\Gamma_{RS}'=  \left( \frac{L_f}{L_w}\right)^{1/4}  \frac {\Gamma_f \sqrt{\sigma_f}}{ \Gamma_w}
\ea

Strong shock conditions require that post-shock temperature are relativistic. For the forward shock
\ba &&
T_{FS} = \frac{1}{8 \sqrt{\sigma_w} }\frac{\Gamma_{FS}}{2 \Gamma_w } > 1 \to \frac{1}{8}  \left( \frac{L_f}{L_w}\right)^{1/4}
\nn && 
L_f \geq 4 \times 10^3 L_w
\label{Lf}
\ea

For the reverse shock
\ba &&
T_{RS} = \frac{1}{8} \left( \frac{L_f}{L_w}\right)^{1/4} \frac{\Gamma_f}{\Gamma_w}
\nn && 
\Gamma_f \geq 8 \left( \frac{L_f}{L_w}\right)^{1/4} \Gamma_w
\ea

Curiously, the  fraction of the flare-wind  $\eta_{RS} $ used to push the magnetar wind increases with its magnetization
\be
\eta_{RS} \sim \frac{1}{\Gamma_{RS} ^2} =4   \left( \frac{L_w}{L_f}\right)^{1/2} \frac{\sigma_f}{\Gamma_w^2}
\ee
This is because for higher $\sigma_f$ the RS moves faster through the flare wind ($\eta_R$ is not the dissipated power  that is put into particles, that one  is smaller by $\Gamma_{RS}'/(8 \sqrt{\sigma_f})$). 

The main constraint for the wind-wind interaction  comes from the fact that the flare wind  should be supersonic with respect to the magnetar wind.
To make a shock in the preceding wind it is required that 
\be
\Gamma _f> \Gamma^{\ast}= 2 \Gamma_w \sqrt{\sigma_w} = 2 \Gamma_w^2
\ee

Since acceleration of the flare proceed according to the same law (\ref{rw}), this requires
\ba &&
\Gamma_f = \sqrt{\sigma_f}
\nn && 
\sigma_f \geq 4 \Gamma_w ^4
\ea
Since terminal magnetization is related to the parameter $\mu_w$ by (\ref{kw2}),
it is required that
\be
\mu_f = \sigma_f ^{3/2} =  8 \Gamma_w^6
\ee

Thus the flare  wind must be much cleaner than the initial  wind
\be
\frac{\mu_f }{\mu_w} >  8 \Gamma_w^4 
\label{muf} 
\ee

\subsection{The fluid engine}
\label{fluid} 
Above we took an extreme position, that the energy transfer from the ejecta to the wind was instantaneous, zero mass {\it and}  zero magnetic loading at the explosion site. In fact, energy transfer between two  relativistically  expanding and relativistically accelerating flows will be  inefficient. Two factors are at play.
First, there  is a problem with creation of shocks during the acceleration stage of the wind: in the acceleration region the magnetization is extremely high, Eq. (\ref{sigma1}). It is very hard to create a shock in a highly magnetized flow.  Consider for example a fluid explosion starting with  ejection at $R_{ej} 
\leq R_{LC}$.
 Acceleration of the fluid ejecta is linear at first \citep{1986ApJ...308L..43P}
\be
\Gamma_{ej} = \frac{r}{R_{ej}}
\ee
At the  wind acceleration stage $r< R_w$, the condition that the ejecta makes a  strong  shock in the preceding wind is (see Eq. (\ref{sigma1})
\be 
\Gamma_{ej} \geq 2 \sqrt{\sigma(r)} \Gamma(r) = 2 \mu_w^{1/2} \sqrt{\frac{r}{r_{LC}}}
\ee
This requires 
\be 
 \frac{r}{R_{LC}} \geq (256 \mu_w)  \left( \frac{R_{ej}}{R_{LC} }  \right) ^2  
\ee
Since $\mu \gg 1$, the  fluid ejecta just starts to make a shock at large distances. 


Second, shock or not, if there is an over-pressurized region with energy density $u$ in the lab frame, which is expanding with \Lf\ $\Gamma$, then the energy density in the  frame associated with the interface  of  the  two interacting media is $\sim u/ \Gamma^2$. This is the force per unit area that contributes to the acceleration of a lower-pressure region. The acceleration time (energy transfer time)  in the lab frame $t_{acc}$  is slower by another factor of $\Gamma$,  hence  the shortest  time to transfer energy from the ejecta to the previous wind is  $t_{acc} \sim  \Gamma^3 (r/c)$. For example 
if interaction starts at $R_w \sim \Gamma_w R_{LC} $ (\ref{rw}), the energy transfer time is $ \sim   \Gamma_w^4  R_{LC} $.

\section{Magnetically-driven explosions}
\label{Magnetically-driven}

\subsection{Not ``magnetic shells"}

Above we summarized the dynamics of magnetized shocks in magnetized winds. We did not address in details the energy release process, just highlighted in \S \ref{fluid} possible issues with the fluid engine. The early dynamics is the key. As we are interested in magnetic explosions from highly magnetized magnetar, the engine and the explosion are  both magnetic.

 Typically magnetic explosions  were  considered in a framework of the internal shock model of GRB \citep{2004RvMP...76.1143P}, as analogue of flying  fluid shells \citep[\eg][]{2020ApJ...896..142B}, but with internal \Bf.   This is not correct  in principle: dynamics of magnetic explosions cannot be reduced to ``magnetized shells".

First, in the case of  {\it one-dimensional} magnetic explosion 
\cite{2010PhRvE..82e6305L} found  a fully analytic {\it one-dimensional}  solution, a simple wave,  for a non-stationary expansion of high magnetized plasma into vacuum and/or low density medium. 
The result is somewhat surprising:   initially the plasma accelerates as $\Gamma \propto t^{1/3}$ and can reach terminal \Lf\ $\Gamma_f=1+2 \mu_0$. 
 where $\mu_0$ is the initial magnetization \citep[see also][]{2010ApJ...720.1490L}.  
 Thus, it was shown, that time-dependent explosions can achieve much larger {\Lf}s that the steady state flows: $2 \mu_0$  for non-steady versus $\mu_0^{1/3} $ for steady-state (in this particular  application $\mu_0 = \sigma_0$).
  The force-free  solution were generalized to the MHD regime \citep[using a mathematically tricky  hodograph transformation  by][]{2012PhRvE..85b6401L}. 
  
  Importantly, those  were  specifically {\it  one-dimensional}  models of local break-out of magnetized jet from, \eg\ a confining star in GRB outflows. They   {\it cannot} be simply used to describe fully three-dimensional magnetic explosions.
What is different in multidimensional magnetic explosions   is the conservation of the   magnetic flux. Such  issues does not appear for one-dimensional motion. 

As \cite{1974MNRAS.167....1R,1984ApJ...283..694K}  argued,  the presence of the global  \Bf\ changes  the dynamics completely if compared with the fluid case. \cite{2003astro.ph.12347L,2006NJPh....8..119L} reformulated the problem in terms of the magnetic flux: the  $\sigma$-problem is the problem of the (properly defined) toroidal magnetic flux conservation.   Below we apply those ideas to the launching region of magnetic explosions.

\subsection{Dynamics of magnetic explosion}
\label{tangled} 
Let's assume that a magnetospheric process (\eg\ flare-like event) created a magnetic bubble, disconnected from the rest of the \ms. 
The bubble   contains both toroidal and poloidal magnetic flux, Fig. \ref{magnetictube}.

 \begin{figure}[h!]
\includegraphics[width=.49\linewidth]{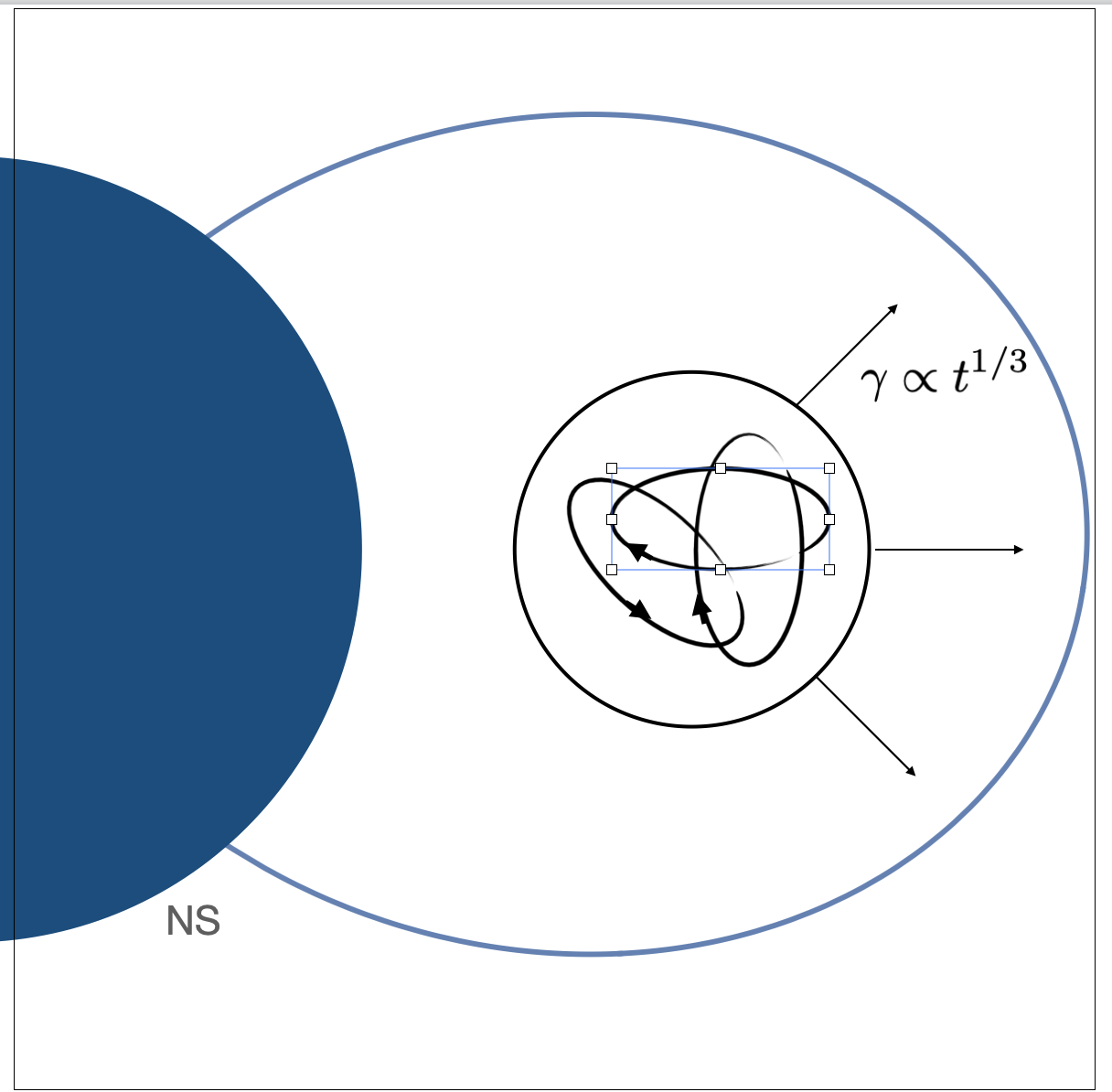}
\includegraphics[width=.49\linewidth]{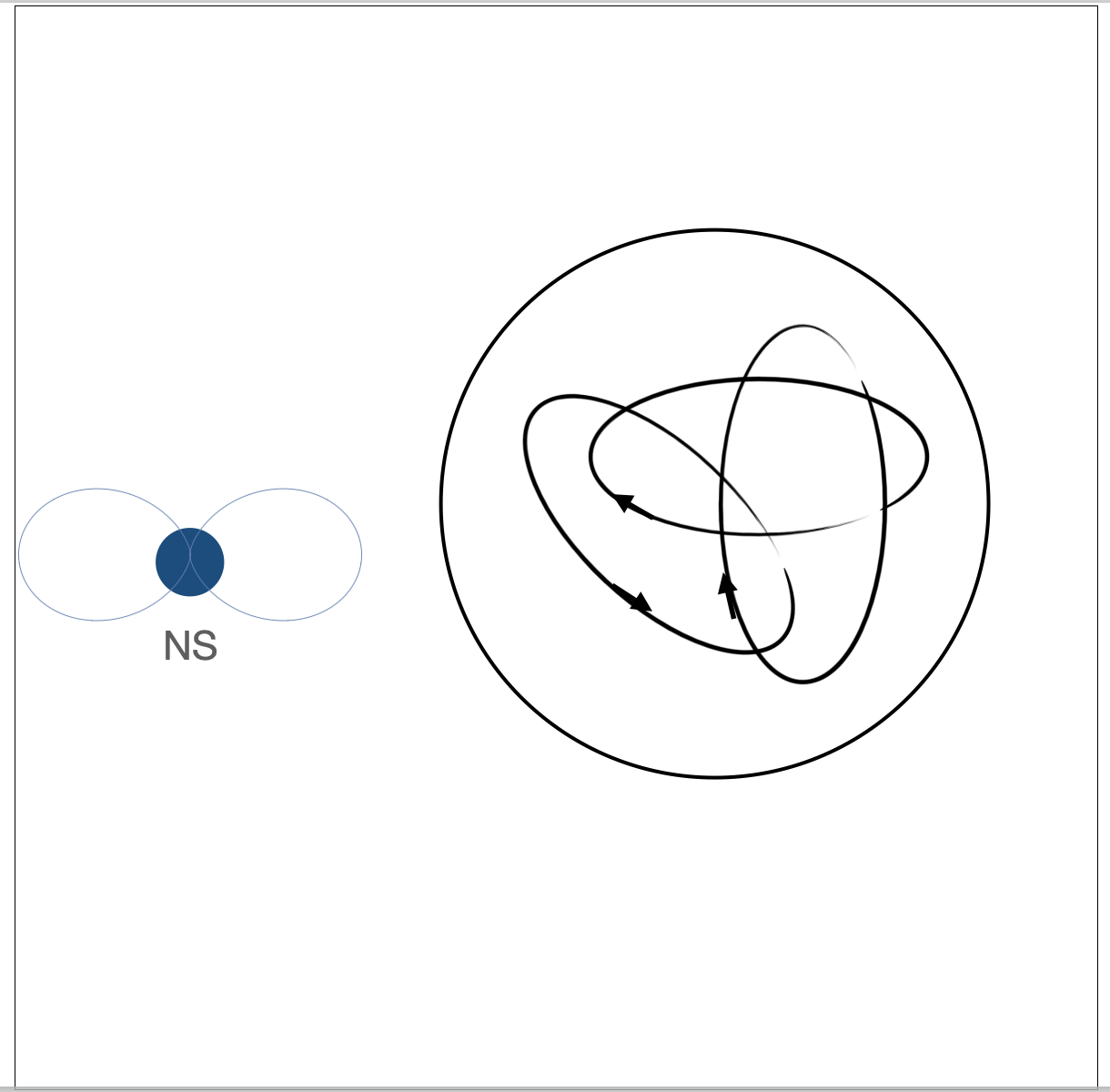}
 \caption{Cartoon of magnetic explosion. Left panel:  a Solar flare-type even near the surface of the \NS\ produces a magnetically  disconnected magnetized cloud  with complicated  (linked)  internal magnetic structure  and total internal  pressure (\Bf\ and pairs)  slightly  exceeding the local dipole field. As the magnetic blob expands its pressure quickly becomes larger than of the surrounding dipolar one. Expansion  quickly becomes relativistic. 
Later on, as the blob  expands,  magnetic field in the blob scales as $B_b \sim 1/R^2$, much faster the \Bf\ in the wind. As a result, the  expelled blob quickly reaches pressure  equipartition with the wind flow. 
  }
\label{magnetictube1}
\end{figure}

The strong gradient of the dipolar field will push the magnetic bubble away.  It will quickly become highly over-pressurized, and will start expanding. 
The expansion will have two stage. First,  near the surface of the bubble  the field is purely tangential. 
At this stage  expansion is described by the 1D model of \cite{2010PhRvE..82e6305L}. If initial magnetization within the bubble was $\sigma_0$, the flow becomes supersonic at the moment the expansion reaches $\Gamma = (\sigma_0/2)^{1/3}$. Unlike the stationary wind when acceleration ends at the sonic point, non-stationary magnetic explosion continues to accelerate to $\Gamma = 1+ 2\sigma_0$.

After a rarefaction wave propagated deep inside the magnetic bubble, the expansion dynamics changes. The bubble consists of closed magnetic loops.
Flux conservation requires  that the \Bf\ with the bubble scales as $B \propto 1/R^2$, where $R$ is a current size of the bubble.   As long as the expelled blob is inside the \LC, the dipolar field  decays much faster $\propto 1/R^3$ (the bubble is also moving away from the star, so it is located approximately at a  distance $R$ comparable to its size), so the expansion of the blob occurs almost like in vacuum. The energy contained within a blob decreases
$E \sim B^2 R^4 \propto R^{-1}$. 
When the size of the expanding bubble becomes larger than  the \LC, the external \Bf\ change its scaling from $\propto R^{-3}$ to   $\propto R^{-1}$. Thus, the bubble quickly reaches equipartition with the wind field.

For example, if the bubble is created near the \NS\ with typical energy
\be
E_f \sim B_f^2 R_f^3
\ee

The \Bf\ inside the newly created magnetic cloud is of the order of the NS's surface field $B_{NS}$,  $B_f \sim B_{NS}$,  and a size smaller than 
$R_{NS}$,  $ R_f =\eta_R R_{NS}$,    $\eta_R \ll 1 $. (Only giant flares need  $\eta_R \sim 1$, FRBs  with nearly quantum \Bf\ need about a football field of energy to account for the high energy emission, $ \eta_R \sim 10^{-2}$, see \S \ref{1935}.)  So,
\be
E_f \sim \eta_R^3 B_{NS} ^2 R_{NS}^3
\label{Ef}
\ee

Given the $\propto 1/R^2$ decrease of the \Bf\ within the bubble, the 
pressure balance between the expanding flux tube and  the wind at $r \geq R_{LC}$ gives the radius $R_{eq} $ when the expanding flux tubes reaches the force balance with the wind 
\be
\frac{R_{eq}}{ R_{LC}} =  \eta_R^{3/2} \frac{ R_{LC}}{R_{NS}}
\label{Req}
\ee
Since the $ \eta_R \ll  1$, this is achieved  fairly close to the \LC. 

At larger distances the bubble is just advected with the wind, always kept at the force balance on the surface. The light bubble is also  first accelerated with the wind,  and then coasts (at $r> R_w$). Since in the coasting stage the confining \Bf\ decreases as $1/r$, the size of the bubble increases as 
$R_{bubble} \propto r^{1/2}$. Eventually, when the magnetar's wind starts interacting with the ISM,  the ejected bubble shocks the ISM and produces radio afterglow    \citep{2005Natur.434.1112C,2005Natur.434.1104G} seen after the giant flare of SGR 1806-20  \citep{2005Natur.434.1107P,2005Natur.434.1098H}, as described by \cite{2021MNRAS.506.6093M}

\subsection{Where did all the energy go? - Magnetic loading}

We seem to run into a little paradox. In the case of FRBs the newly created magnetic bubble with energy (\ref{Ef}) 
would have more energy than  the energy of the \Bf\ measured at the \LC, within the volume of a \LC,
\be
E_f \gg E_{LC} \sim \frac {B_{NS}^2 R_{NS}^6}{R_{LC}^3}
\ee
Yet, the expanding bubble came into force-balance with the preceding wind near the light cylinder. 
 Where did the  extra energy go?
 
  It went  into  the stretching of the  internal \Bf\ of the bubble.  Recall that  the pressure of the \Bf\ {\it along}  the field is negative: the fields ``wants'' to contract.  Stretching the field (expanding  loops of the tangled  \Bf) requires work to be done against the contracting parallel force. The over-pressurized magnetically-dominated  configuration ``forces''  the field to expand, thus making work on the  internal field. 
As a result, most of the excess magnetic energy is spent on stretching the internal \Bf, not on producing shocks/making $pdV$ work on the surrounding medium. 
This  effect can be called  a  magnetic loading.

We come to an important conclusion: magnetic explosions are dominated not by the mass loading, but by the magnetic loading. Even very powerful explosion, with energies much larger than $B_{LC}^2 R_{LC}^3$ reach a force balance close to the \LC.  After that they are locked in the flow, and quietly advected.

\subsection{Expanding spheromak/flux ropes}

\begin{figure}[h!]
\includegraphics[width=.49\linewidth]{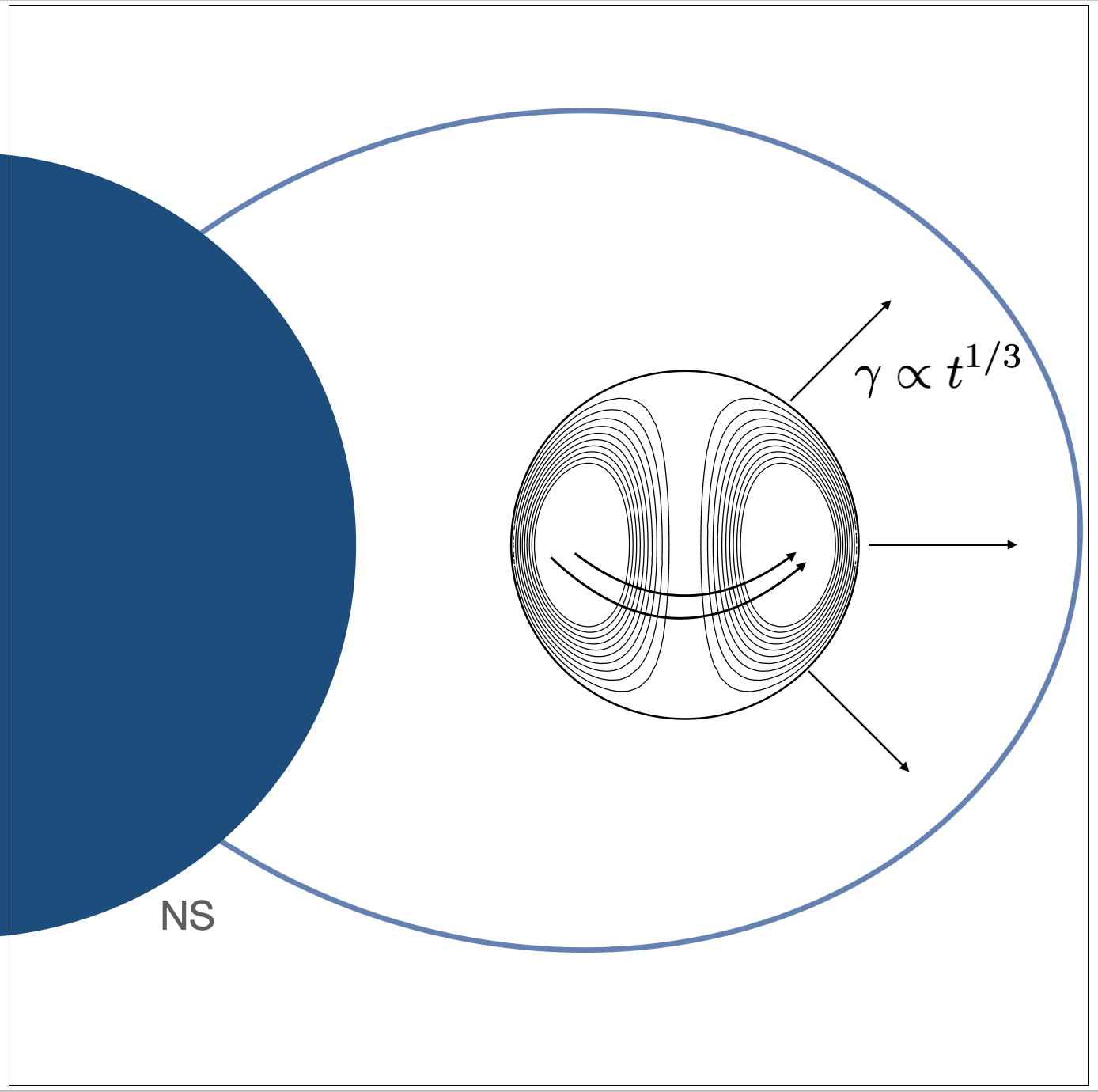}
\includegraphics[width=.49\linewidth]{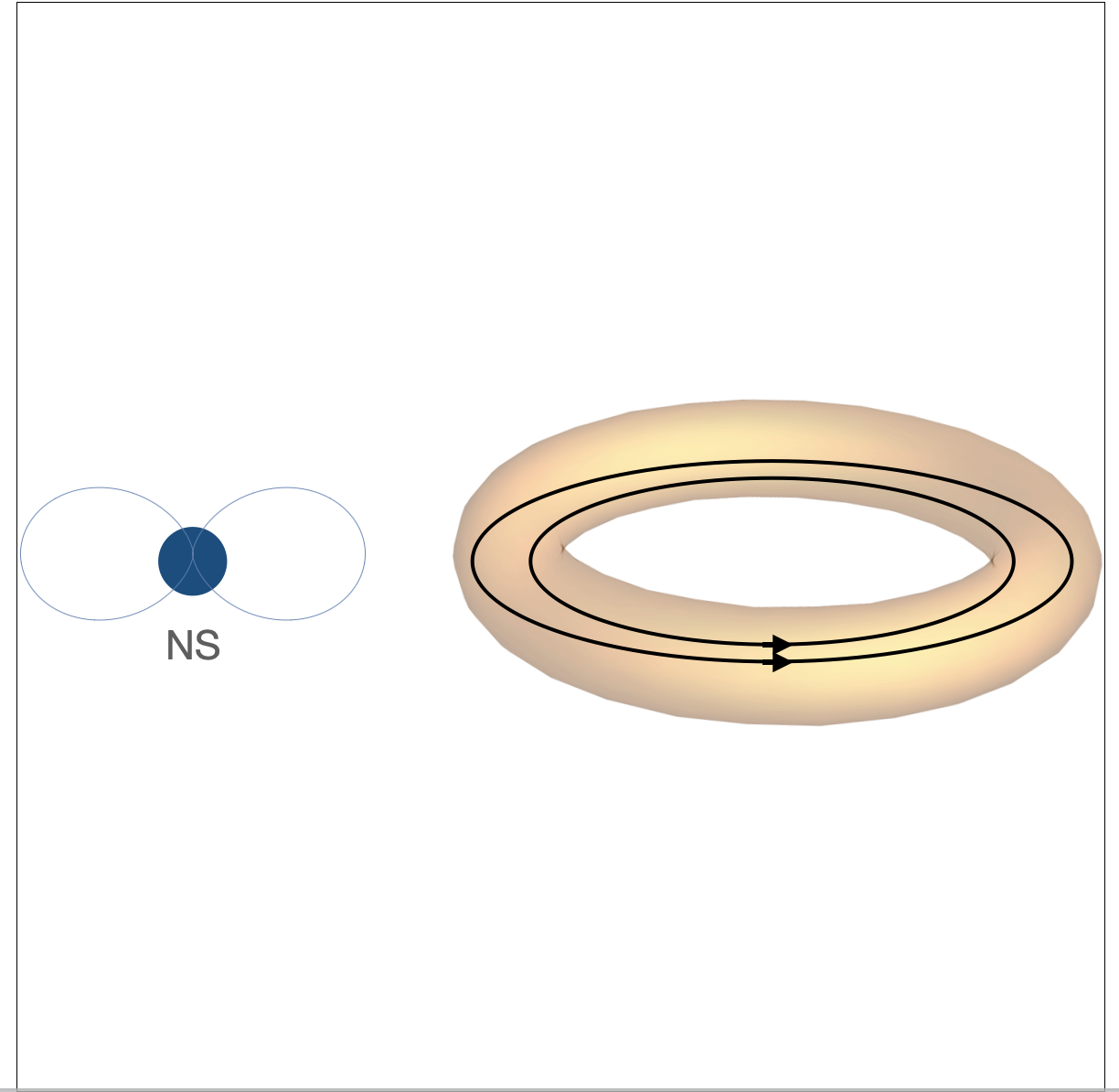}
  \caption{Cartoon of magnetic explosion of a spheromak. Left panel:  a Solar flare-type even near the surface of the \NS\ produces a magnetically  disconnected magnetized cloud, a spheromak. Expansion is mostly equatorial.  Right panel:  During further expansion the magnetic flux in the tube remains constant. As a result the tube quickly reaches the force-balance with the wind. After that the flux tube forms a conically expanding structure.  }
\label{magnetictube}
\end{figure}

To illustrate the above points with  the fully analytical model, and suggest possible variations, one can assume that a newly created bubble resembles a spheromak \cite{BellanSpheromak}. 
A possible model of the newly created magnetic cloud is a  slightly over-pressured spheromak, Fig. \ref{magnetictube}.

We can appeal  then to the fully analytical solution 
for  non-relativistically  expanding spheromak  \cite{2011SoPh..270..537L}:
 \ba &&
 B_r =  2 B_0  \alpha { j_1 \over \alpha_0^2 r} \cos \theta, 
\nn &&
B_\theta =-B_0 \alpha  { j_1 + \alpha r j_1' \over \alpha_0^2 r} \sin \theta,
\nn &&
B_\phi = B_0 j_1 \left({\alpha\over \alpha_0}\right)^2  \sin \theta,
\nn &&
E_r=0,
\nn &&
E_\theta= - B_0  { \alpha \dot{\alpha} \over  \alpha_0^2}  r j_1 \sin \theta=- B_\phi {\dot{\alpha} \over \alpha} r,
\nn &&
E_\phi =- B_0  \dot{\alpha}  { j_1 + \alpha r j_1' \over \alpha_0^2 } \sin \theta= B_\theta {\dot{\alpha} \over \alpha} r,
\nn &&
{\bf E}=   \left( {\dot{\alpha} \over \alpha} r  \right) {\bf e}_{r} \times {\bf B}.
\label{main}
 \ea
where $j_1$ is a Bessel function, $\alpha_0 \sim 1/R_0$ is the initial radius, $\alpha \sim 1/R$ is the current radius. 
The total magnetic helicity, $ {\cal H} = \int dV {\bf B} \cdot {\bf A}=1.4 \times 10^{-2} B_0^2 R_ 0^4$, as well as  the toroidal magnetic flux 
${\cal F} =   \int  B_\phi r dr= 5.26 B_0 R_0^2$  are  constant. Importantly, note: $B \propto 1/R^2$.

The  total energy of the expanding spheromak
\be
E_{\rm tot} = \int d V { E^2 + B^2 \over 8 \pi}  \propto  { B_0^2 R_0^4 \over R}
\ee
decreases. 
This is a fully analytical solution illustrating the principal points made in \S  \ref{tangled}.


The late expansion of a spheromak will be somewhat different though. On the surface the \Bf\ of the spheromak scales as  $B_\perp \propto \sin \theta$. Thus, initial expansion will be mostly equatorial. The expanding  part of the spheromak,   with high toroidal \Bf,  becomes causally disconnected and forms an expanding magnetic flux tube,   right panel in Fig. \ref{magnetictube} \citep[][also discussed the expanding flux rope, though in that case the solution is approximate]{2011SoPh..270..537L}.

Some difference in evolution between an expanding bubble and a flux tube  appears after they reached force-balance with the wind. In case of the flux tube, 
consider a magnetic flux tube initially   containing  \Bf\ $B_0$, overall radius $R_0$, radial extent $\Delta R$ and polar angle extent of $ \Delta \theta$. 
It is expected that during expansion $\Delta R$   and  $ \Delta \theta$ remain fixed.
Conservation of the magnetic flux 
$\Phi  \approx B (\Delta \theta) r (\Delta R)$ then implies
\be
B= \frac{R_0}{r} B_0
\ee
This matches the \Bf\ in the wind. Thus, after reaching a force balance close to the \LC\ the ejected flux tube remains in force-balance with the wind. The energy contained in the flux tube remains constant
\be
E= B^2 (\Delta \theta) r^2 (\Delta r)=E_0
\ee
Thus,  expanding magnetic flux tube does not do any work on the surrounding.

In passing we note that both spheromaks and flux tubes models were discussed for  Solar CME \citep{Farrugia95}.

\subsection{Force-free pulse}
\label{Michles}

Let us complement  the above description  with a related  fully analytical non-linear  model of temporarily  spun-up \NS, a jerk in a spin. The model discussed below is a variant of the magnetospheric ejection model, approximating  a flare induced by crustal  motion. The difference with the blob case is that now  it is not assumed that an isolated magnetic bubble/flux tube is generated. Yet though the set-up is different, the conclusion is the same as above: magnetic perturbations do not generates shocks  (but propagating EM pulses in this case).

Qualitatively, "a jerk" in the angular rotation velocity $\Omega$  mimics a shearing motion of a patch of field lines \citep[in fact, a star quake, often invoked as a driving mechanisms of flares, was shown to be a problematic model of flares by][]{2012MNRAS.427.1574L}.
The analytical model described below   misses the magnetospheric dynamics, but it capture the wind dynamics.

\begin{figure}[h!]
\includegraphics[width=.33\linewidth]{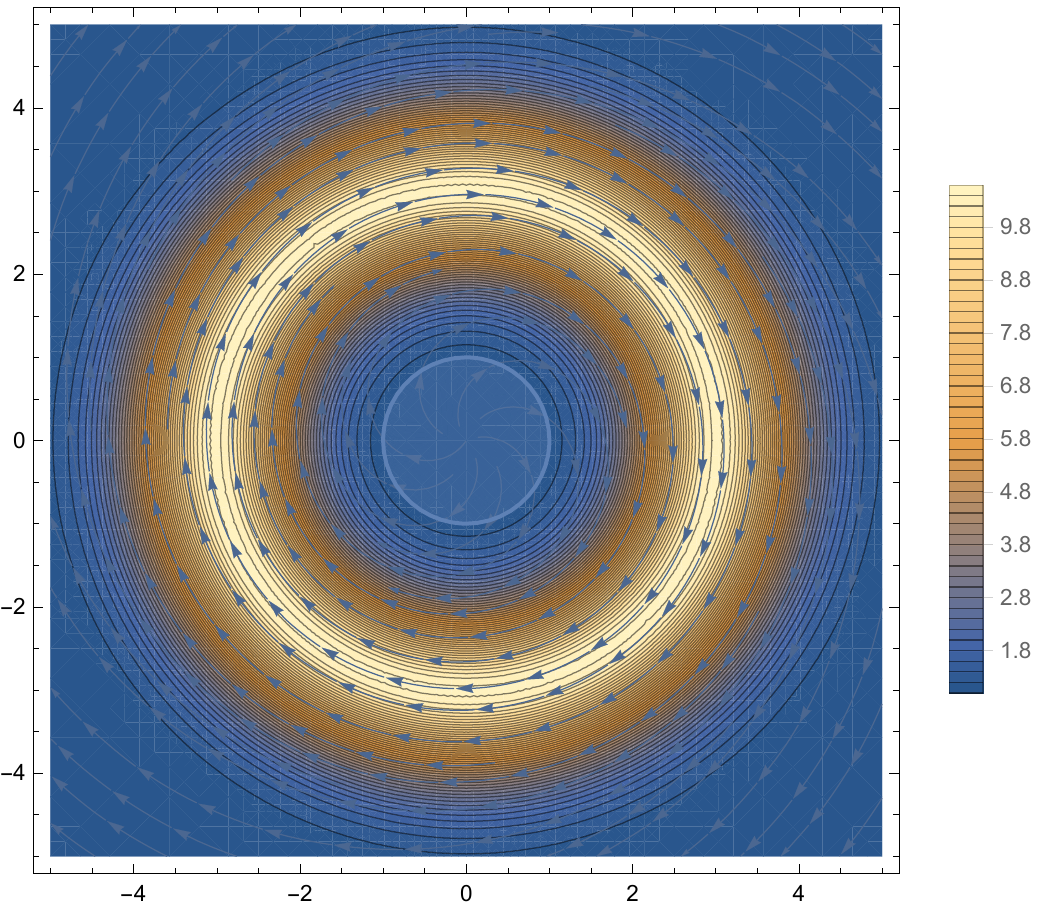}
\includegraphics[width=.33\linewidth]{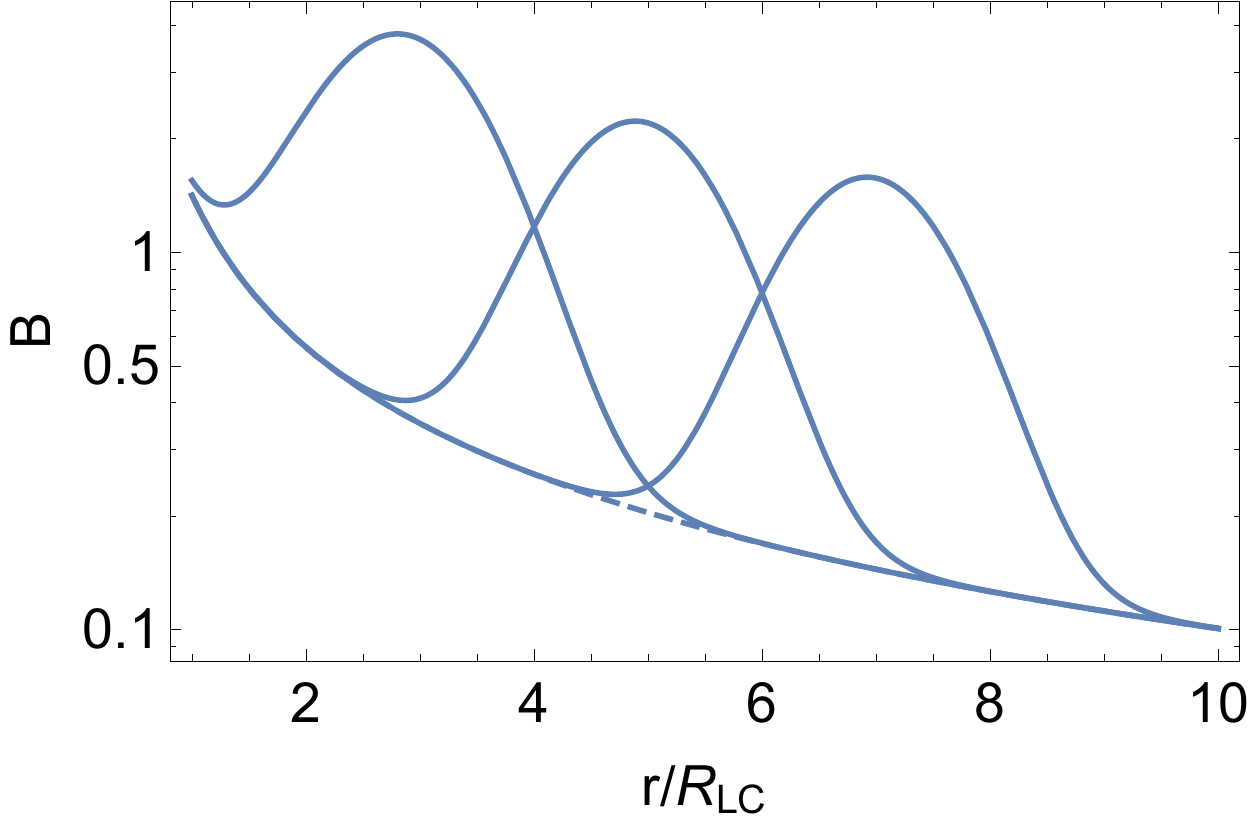}
\includegraphics[width=.33\linewidth]{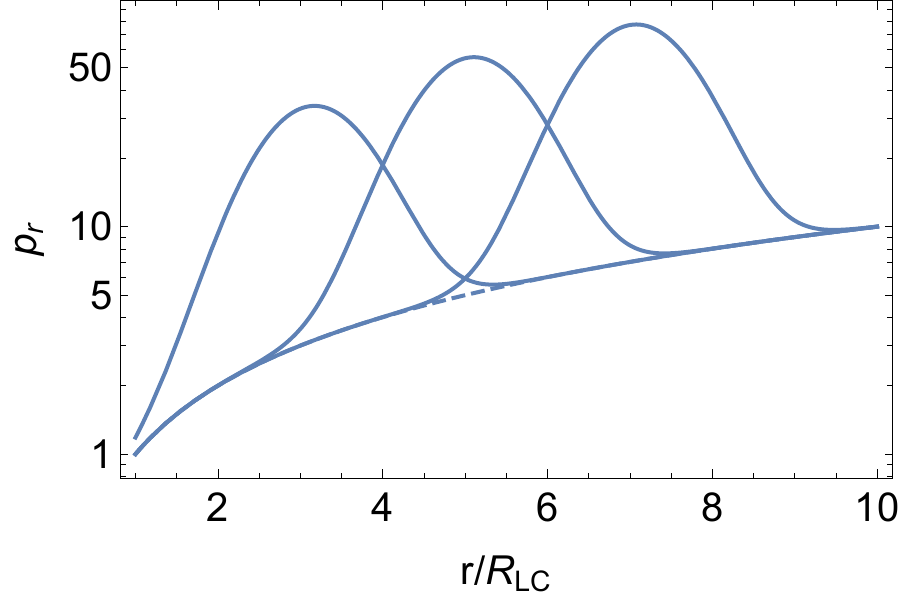}
\caption{Left Panel: Structure of the \Bf\ in the equatorial plane of an \EM\ Gaussian  pulse with amplitude $10$ times the average propagating through Michel's wind. The peak of the pulse is at $r/R_{LC}=3$. The \LC\  is at $\sqrt{x^2+y^2}=1$. Color scheme corresponds to $\ln B/B_M$, where $B_M$ is the local value of the magnetic  field for Michel's solution. Center Panel: plot of $ B(r)$ showing EM pulse propagating with the wind for times $t=3,5,7$ (in units of $R_{LC}/c$); dashed line is the Michels' solution.  Right panel: plot of $ p_r(r)$. The pulse propagates with the flow with constant relative amplitude,  without experiencing any distortions.
}
\label{pulseff}
\end {figure}

\begin{figure}[h!]
\includegraphics[width=.33\linewidth]{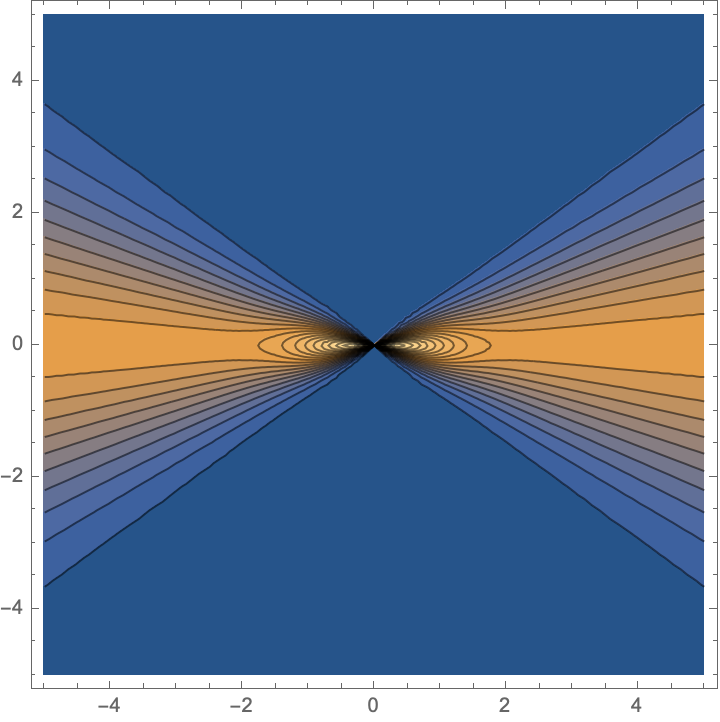}
\includegraphics[width=.33\linewidth]{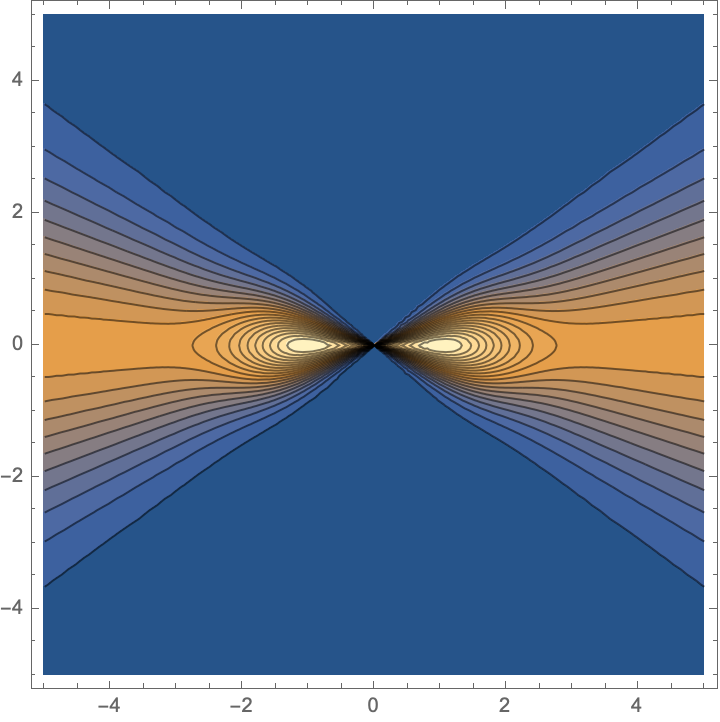}
\includegraphics[width=.33\linewidth]{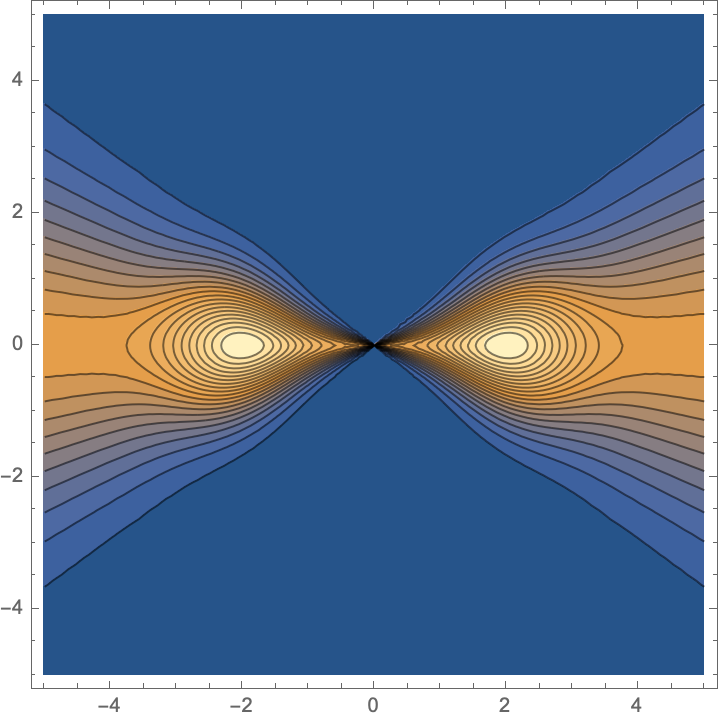}\\
\includegraphics[width=.33\linewidth]{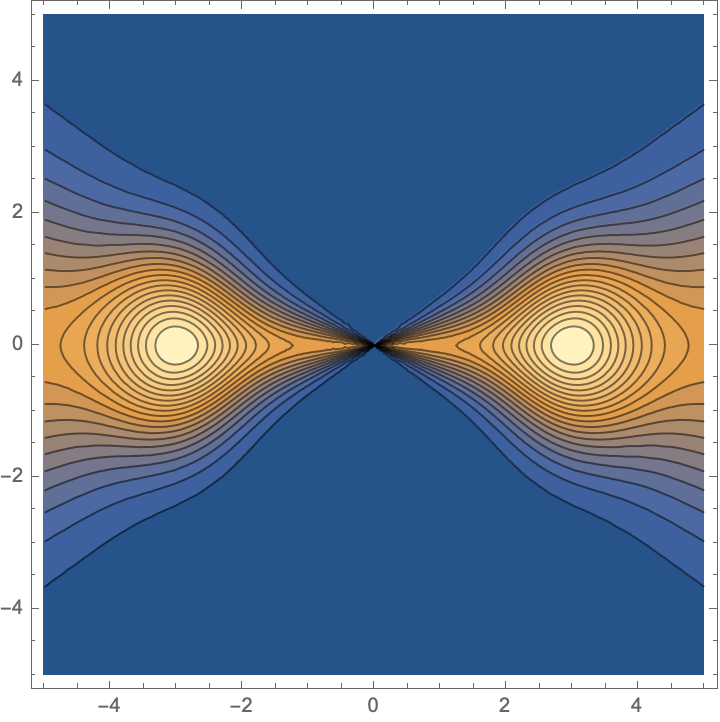}
\includegraphics[width=.33\linewidth]{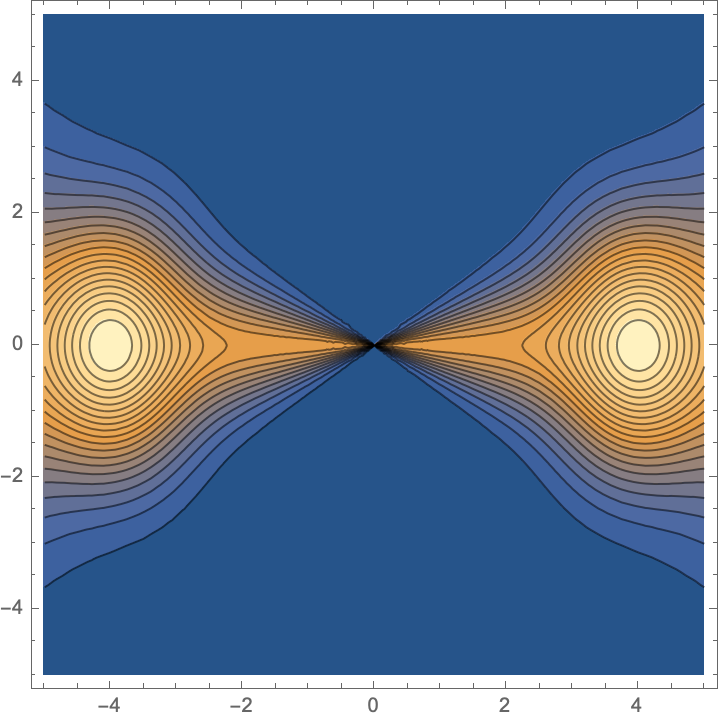}
\includegraphics[width=.33\linewidth]{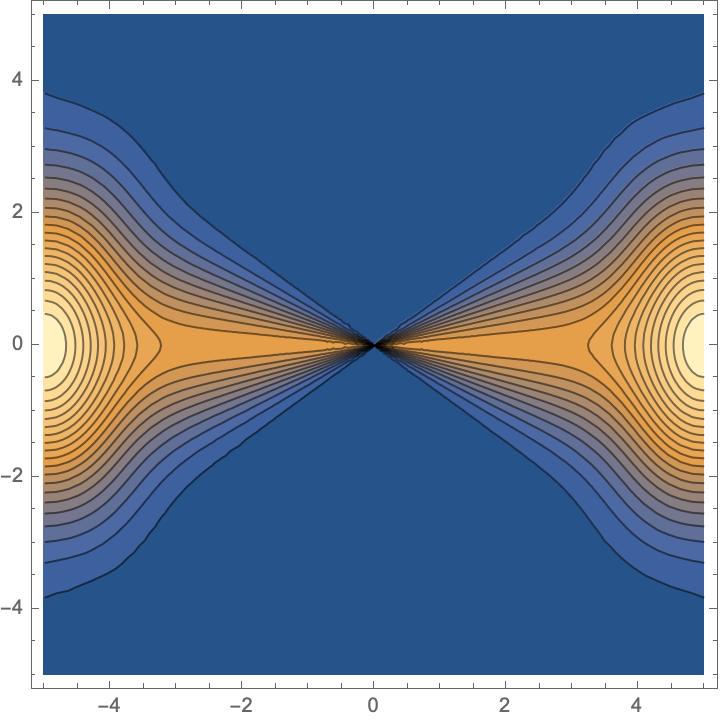}
\caption{Dynamics of ejected flux tube with amplitude of $1$. Color plot is value of $B_\phi$. An ejection is parametrized as  $g(\theta)= \sin^{10} \theta $, $\Omega = 1+ e^{(r-t)^2}$.  Times are $t=0,1,...5$. These are fully analytical solutions, Eq.  (\protect\ref{FFF}).
}
\label{TTshort}
\end {figure}

If the magnetization of the wind is sufficiently high, we can actually build {\it exact time-and-angle dependent model of propagation of the force-free \EM\ pulse}. 
As shown by \cite{2011PhRvD..83l4035L} \citep[see also][]{2014MNRAS.445.2500G} a stationary solution of monopolar \ms\ by \cite{1973ApJ...180L.133M} can be generalized to arbitrary $\Omega=\Omega  [r-t] g( \theta)$:
  \ba &&
  B_r = \left( {R_{\rm LC} \over r} \right)^2 B_s
  \nn &&
   \, B_\phi =  { R_{\rm LC}^2  \sin \theta \over r} B_s  \Omega  (r-t) g( \theta)
   \nn &&
    \,  E_\theta = B_\phi
 \label{FFF}
\ea
see Fig \ref{pulseff}- \ref{TTshort}.
 This  time-dependent nonlinear solution (nonlinear in a sense that  the current is a  non-linear  function of the magnetic flux function) preserves both the radial and $\theta$ force balance. 
It can also be generalized to \Sc\ metric using the  Eddington-Finkelstein coordinates \citep{2011PhRvD..83l4035L}.

A pulse with $\Omega [r-t] g(\theta) $  propagates  with  radial 4-momentum
\be
p_r =   \frac{ r^2  \sin^2 \theta(  \Omega  [r-t] )^2 g(\theta)^2}{ \sqrt{ 1+  r^2  \sin^2 \theta(  \Omega  [r-t] )^2 g(\theta)^2}}
\ee
which is larger than that of the wind for $ \Omega  [r-t]  2 g(\theta) \geq \Omega_0$, the constant value.
Higher radial  momentum of the ejected shell  (than of the background flow)  does not mean that plasma is swept-up: it's just an EM pulse propagating through the accelerating  wind.

The solutions  (\ref{FFF}) incorporate an EM pulse with a shape of  arbitrary radial and angular dependence. For example, choosing a Gaussian pulse with  $\Omega= ( 1 + \delta e^{-(r-t)^2/(\Delta r)^2}) \Omega_0$, 
solutions (\ref{FFF}) represent a pulse with total power $
\frac{1}{4} \sqrt{\pi } \delta  \left(\sqrt{2} \delta +4\right) \approx \delta^2$
larger than the average wind power. 

Since realistic dipolar \mss\ do evolve asymptotically to the  \cite{1973ApJ...180L.133M} solution \citep{1999A&A...349.1017B,2006MNRAS.367...19K,2009MNRAS.394.1182K}, we conclude that   a force-free pulse with  arbitrary amplitude and shape   does not evolve (neither in $\theta$  nor $r-t$ coordinates) as the flow expands and accelerates.

This analytical example differs somewhat from the ejected blob/flux tube case in that it does not satisfy the ``zero normal component of the \Bf\ at the surface of the blob. Hence the  importance of this example: magnetic ejection are either advected with the wind, or propagate as non-dissipative EM  pulses. No shocks are generated.

The model of a "jerk in spin" described above has a limitation that inside the \ms\ the corresponding \EM\ pulse, produced as presumed by the sudden crustal motion,  is expected to break well before it reaches the \LC\ \citep{1996ApJ...473..322T}. As the crustal motion-generated    EM  pulse propagates through the \ms,  the relative  amplitude of the O-mode  scales as $\propto r^{3/2}$. This will lead to wave breaking  \citep{1975OISNP...1.....A,1974PhRvL..33..209M,1974PhFl...17..778D}  and dissipation of the wave energy \citep[in simulations of ][ wave breaking was artificially prohibited by resetting the \Ef\ to $E\leq B$ at every time step]{2020ApJ...900L..21Y}

\subsection{Applications to SGR 1935+2154 X-ray/radio flares}
\label{1935}

In the case of SGR 1935+2154, 
the unambiguously associated with a short pulse   Burst-G by classification of \cite{2020ApJ...898L..29M},  had peak X-ray  luminosity of $\sim 10^{40}$ erg/s and total released energy $E_f \sim 10^{39}$ erg. The G burst  was also particularly spectrally hard. The accompanying radio burst had $3 \times 10^{34}$ ergs; the radio to high energy fluence was $F_R/F_X \sim 2 \times 10^{-5}$.  The period is $3.4$ seconds (so $R_{LC} = 1.4 \times 10^{10} $ cm,  $R_{LC}/R_{NS} = 1.4 \times 10^4$). 
The spindown luminosity is  $1.7 \times 10^{34} $ erg s$^{-1}$, the surface \Bf\ is $B_{NS} = 2.2 \times 10^{14}$ G \citep{2016MNRAS.457.3448I}

Since the process of launching of any outflow is energetically the most demanding step, and assuming that a $\sim$ half of the energy was dissipated, we estimate the ejection energy as the energy of the X-ray burst.
The required volume of the \ms\ that got dissipated to power the X-ray flare is
\ba &&
R_f \sim \frac{E_f^{1/3}}{B_{NS}^{2/3}}= 3\times 10^3\, {\rm cm}
\nn && 
\eta_R = \frac{R_f}{R_{NS}} = 2\times 10^{-3}
\ea
A flare  was just only 30 meters in size.

Expected wind coasting radius (\ref{rw}) is  $R_w= 1.7 \times 10^{12} (\Gamma_w/100)$ cm, 
scale    (\ref{R00}) is $R_0 = 1.7 \times 10^{15} $ cm. 

The flare released energy much larger than the magnetic  energy at the \LC,
\be
\frac{E_f }{E_{LC}} = 6\times 10^4
\ee
The equipartition radius (\ref{Req}) evaluates to
\be
\frac{R_{eq}}{ R_{LC}} = 2
\label{Req1}
\ee
Thus the expanding blob reached equipartition right near the \LC, consistent with our  conclusion on the general FRB population.

Another problem for  the wind models comes from fairly small compactness parameter.  At peak luminosity of $\sim 10^{40}$ erg/s   the compactness parameter
$l_c$ evaluates to 
\be
l_c =\frac{\sigma_T L_\gamma }{m_e c^4 \tau _\gamma}\leq 10^3
\ee
(actually smaller since the peak energy of the burst was $E_p \sim $ 65 keV$\ll m_e c^2$). Omitting  for the sake of an argument the magnetic loading, 
the corresponding hydrodynamic  terminal   \Lf\ is small
\be
\Gamma_f \propto l_c^{1/6} \sim \mbox{few}
\ee
for  duration of  $\tau _\gamma \sim 10$ msec.

\section{Discussion}
\label{Discussion}

We demonstrate that  the wind models  of FRBs are internally inconsistent on several grounds. The wind-type models of FBRs had  a clear ``good'' point:  {\it if}  (only  {\it if} !) one can make a  relativistic shock in a relativistically  receding wind, the combined \Lf\ of the shock  and (two times)  the wind's \Lf\ make  for an extremely high \Lf\ shock, hence short durations from large distances.

There  are serious  problems with this scenario. First, regardless of the launching mechanism, the flare-generated outflow must be 
exceptionally clean, launching an outflow with the \Lf\ {\it at least} few thousands,  Eq. (\ref{kw2}) assuming very low $\Gamma_w \sim 30$. This is about an order of magnitude higher than in GRBs.
Wind models of the ejection are hopeless, the ejecta-wind must be millions of times cleaner than the preceding magnetar wind, Eq. (\ref{muf}). Impulsive ejections have an advantage that they can reach terminal {\Lf}s $ \sim \mu_0$,  compared with $\mu_0^{1/3}$ for steady state. 

 Most importantly, we argue that in  contrast to the hydrodynamic explosions (when the terminal \Lf\ is determined either by ion loading, or by pair freeze-out),
 the dynamics of the magnetic explosions  is controlled by magnetic loading: the requirement  of the conservation of the  magnetic flux.  Thus, in the case of magnetar explosions the dynamics is  drastically different from the fluid case; instead of shocks with {\Lf}s close to a million, the explosions  reach force-balance near the \LC\ and then either  are advected with the wind   as non-dissipative EM structure, or propagate as EM pulses.




Historically, 
the wind-type models of FBRs started, perhaps,  with the \cite{2014MNRAS.442L...9L}  model. He advocated an EM pulse propagating in the wind (which is a correct model in our view), but the energetics of his model was $ \geq 10^{50}$ ergs, in the ball park of GRBs, and clearly inconsistent with the FRB phenomenology.  Finally we note that the original  cyclotron instability model \citep{hoshino_91,gallant_92,hoshino_92} was developed to explain months long variations of Crab's wisps, some ten orders of magnitude in time scale from the FRBs.




 This work had been supported by 
NASA grants 80NSSC17K0757 and 80NSSC20K0910,   NSF grants 1903332 and  1908590.
I would like to thank Andrei Beloborodov, Yuri Lyubarsky and  Lorenzo Sironi for discussions. 

\section{Data availability}
The data underlying this article will be shared on reasonable request to the corresponding author.

\bibliographystyle{apj}
\bibliography{/Users/maxim/Home/Research/BibTex}

\end{document}